
\def\ds{\displaystyle}
\def\bx{{\bf x}}
\def\by{{\bf y}}
\def\br{{\bf r}}
\def\bp{{\bf p}}
\def\bq{{\bf q}}
\def\kem{{{1\over m} \nabla^2}}
\def\kemm{{{1\over {2m}} \nabla^2}}
\def\phid{\phi^{\dag}}

\def\di{\partial}
\def\ni{\noindent}

\def\phib{\overline \phi}

\magnification=\magstep1
\baselineskip=25pt
\centerline{Exact solutions and `triviality' of $\lambda (\varphi^*\varphi)^2$ theory}
\centerline{ in the Feshbach-Villars formulation}
\par\vskip 2truecm
\centerline{Jurij W. Darewych$^{\dag}$}
\par
\centerline{Department of Physics and Astronomy}
\par
\centerline{York University}
\par
\centerline{North York, Ontario}
\par
\centerline{M3J 1P3}
\par
\centerline{Canada}
\par
\vskip 2truecm

{\ni \bf Abstract}
\par\medskip
The complex scalar (Klein-Gordon) QFT with a $\lambda (\varphi^*\varphi)^2$ interaction is considered in the Feshbach-Villars formulation. It is shown that exact few-particle eigenstates of the QFT Hamiltonian can be obtained.  The resulting relativistic few-body equations  correspond to Klein-Gordon particles interacting via   delta-function, or ``contact'', potentials. Momentum-space solutions of the two-body equation yield a `trivial' unity $S$-matrix. 

\vskip 1truecm
$^{\dag}$darewych@yorku.ca

\vfill\eject

{\ni \bf 1. Introduction}

The model scalar field theory, based on the Lagrangian ($\hbar = c = 1$)
$$
{\cal L}_{KG}=\di^\mu \varphi^*(x) \di_\mu \varphi (x) - m^2 \varphi^*(x)  \varphi (x) - \lambda (\varphi^*(x)  \varphi (x))
^2, \eqno (1)
$$
is often used as a proptotype QFT in many texts (e.g. Gross, ref. [1]), and has been the subject of investigation on its own account in many studies.  To our knowledge, no exact solutions of this theory have been written down.  We shall consider this theory in the present paper, but in the Feshbach-Villars (FV)  formulation [2] of covariant scalar (Klein-Gordon) field theory.
We recall that in the FV formulation, the field $\varphi$ and its time-derivative $\dot \varphi$ are replaced by a two-component vector
$$
\phi=\left[\matrix{u = {1\over {\sqrt{2m}}}(m \varphi + i \dot \varphi)&\cr
                v = {1\over {\sqrt{2m}}}(m \varphi - i \dot \varphi)&\cr}\right], \eqno (2)
$$
so that, for example, $2 m \varphi^* \varphi = (u^* + v^*)(u+v) = \phi^{\dag} \eta \tau \phi$, where $\eta$ and $\tau$ are the matrices
$$\eta = \left [\matrix{1&0\cr 0& -1\cr}\right ]\;\;\;\; {\rm and} \;\;\;\;\tau = \left [\matrix{1&1\cr -1& -1\cr}\right ] \eqno (3).
$$

The Euler-Lagrange equations of motion for this theory are  
$$
\di^\mu \di_\mu \varphi + m^2 \varphi + 2 \lambda (\varphi^* \varphi) \varphi = 0 \eqno (4)
$$
and its conjugate. In the FV formulation the equation of motion (4) takes on the form
$$
i \dot \phi = - {1\over {2m}} \nabla^2 \tau \phi + m \eta \phi + {\lambda \over {2m^2}} (\phib \phi) \tau \phi, \eqno (5)
$$
where $\phib = \phi^{\dag} \eta \tau$.
The  Lagrangian density corresponding to eq. (5) is
$$
{\cal L}_{FV} = i \phi^{\dag} \eta \dot \phi  - {1\over {2m}} \nabla \phib \cdot \nabla \phi - m \phi^{\dag} \phi - {\lambda \over {4m^2}} (\phib \phi)^2, \eqno (6)
$$

Note that ${\cal L}_{KG}$ is not identical to ${\cal L}_{FV}$.  Indeed ${\cal L}_{KG} = {\cal L}_{FV} + {\di \over {\di t}} (\varphi^*{\dot \varphi})$. However, they lead to identical equations of motion (eqs. (4) and (5)) and so are equivalent in this sense. Henceforth we shall base our results on ${\cal L}_{FV}$.
The momenta corresponding to $u$ and $v$ are 
$$
p_u = {\di {\cal L}_{FV} \over {\di \dot u}} = i u^*,\;\;\; {\rm and} \;\;\; p_v = -i v^*,$$
that is, $u^*$ and $v^*$ are, in essence, the conjugate momenta, so that the Hamiltonian density is given by the expression
$$
{\cal H}(x) = \phi^{\dag}(x) \eta {\hat h} (x) \phi(x) +{\lambda \over {4m^2}}(\phib (x) \phi(x))^2, \eqno (7)
$$
where ${\hat h}(x) = \tau (-{1\over {2m}})\nabla^2 + m \,\eta$, and where we have suppressed the term $\nabla \cdot ({1\over {2m}} \phib \nabla \phi)$.

We use canonical equal-time quantization, whereupon the non-vanishing commutation relations are
$$
 [u(\bx,t),p_u(\by,t)] = i \delta^3(\bx-\by)\;\;\;\; {\rm and} \;\;\;\;
[v(\bx,t),p_v(\by,t)] = i \delta^3(\bx-\by), \eqno (8)
$$
or, equivalently,  
$$
[\phi_a(\bx,t),\phi^{\dag}_b(\by,t)] = \eta_{ab}\delta^3(\bx-\by),\;\; \;\;\; a,b=1,2\eqno (9)
$$
and where $\phi^T = [\phi_1=u, \phi_2=v]$, while $\eta_{ab}$ are elements of the $\eta$ matrix (3).
Using these commutation relations, we note that the QFTheoretic Hamiltonian can be written as
$$H = H_0 + H_\lambda, \eqno (10)$$
where
$$
H_0 = \int d^3x \,\phi^{\dag}(x) \eta {\hat h} (x) \phi(x), \eqno (11)
$$
and
$$H_\lambda = {\lambda \over {4m^2}}\int d^3x\,(\phib (x) \phi(x))^2 = {\lambda \over {4m^2}}\int d^3x\,\phib (x)(\phib (x) \phi(x))\phi(x), \eqno (12)
$$
and where we have used $\tau^2 = 0$ in the last step of (12). Note that no infinities are dropped upon normal ordering, since none arise on account of the $\tau^2 = 0$  property.

In the Schr\"odinger picture we can take $t=0$. Therefore, we shall use the notation that, say  $\phi(\bx,t=0) = \phi(\bx)$, etc., for QFT operators.
We define an empty vacuum state, $|\tilde 0\rangle$, such that 
$$
\phi_a |\tilde 0\rangle = 0.  \eqno (13)
$$
This is different from the conventional Dirac vacuum $|0\rangle$ (the ``filled negative energy sea" vacuum), which is annihilated by only the positive frequency part of $\varphi$ and by the negative frequency part of $\varphi^*$ (see, for example, ref. [3], p. 38).  With the definition (13), the state defined as
$$
|\psi_1\rangle = \int d^3x\,\phid (\bx) \eta f(\bx) |\tilde 0\rangle, \eqno (14)
$$
where f(\bx) is a two-component vector, is an eigenstate of the QFT Hamiltonian (10) with eigenvalue $E_1$ provided that the $f(\bx)$ is a solution of the equation
$$
{\hat h}(\bx) f(\bx) = E_1 f(\bx). \eqno (15)
$$
This is just the free-particle Klein-Gordon equation  for stationary states ($|\psi_1\rangle$ is insensitive to $H_\lambda$), in the FV formulation. It has, of course, all the usual negative-energy ``pathologies'' of the KG equation. 
The presence of negative-energy solutions is a consequence of the choice of vacuum (13).
We shall refer to $|\psi_1\rangle$ as a one-KG-particle state.

We can define analogous two-KG-particle states,
$$
|\psi_2\rangle = \int d^3x\,d^3y\; F_{a\,b}(\bx,\by)\;\phid_a (\bx) \phid_b(\by) |\tilde 0\rangle, \eqno (16)
$$
where summation on repeated indices $a$ and $b$ is implied. This state is an eigestate of the QFT Hamiltonian (10) provided that the $2 \times 2$  coefficient matrix $F = [F_{a\,b}]$ is a solution of the two-body equation,
$$\eqalignno{
{\tilde h}(\bx) F(\bx,\by)  + [{\tilde h}(\by) F^T(\bx,\by)]^T  
&+ V(\bx-\by) {\tilde \tau} F(\bx,\by) \tau\cr 
&= E_2 F(\bx,\by),&  (17)
\cr}$$
where the superscript $T$ stands for ``transpose'', ${\tilde h} = \eta {\hat h} \eta$, ${\tilde \tau} = \eta \tau \eta = \tau^T$ and
$$
V(\bx-\by) = {\lambda \over {2m^2}} \delta^3(\bx-\by) \eqno (18)
$$
in this case.  Note that (16) implies that $F_{ab}(\bx,\by) = F_{ba}(\by,\bx)$, or $F^T(\bx,\by) = F(\by,\bx)$ in matrix notation.

Equation (17) is a relativistic two-body Klein-Gordon-Feshbach-Villars-like equation, with a repulsive delta-function interparticle interaction.
If $V =0$, then eq. (17) has the solution $F(\bx,\by)=g_1(\bx) g_2^T(\by)$, where each $f_i(\bx)=\eta\,g_i(\bx)$ is a solution of the free KG equation (15) with eigenenergy $\varepsilon_i$, and where $E_2 = \varepsilon_1 + \varepsilon_2$, as would be expected.
 In the rest frame, ${\bf P}_{\rm total} |\psi_2\rangle = 0$, this equation (17) simplifies to
$$
{\tilde h}(\br) F(\br) + [{\tilde h}(\br) F^T(\br)]^T + V(\br) \tau^T F(\br) \tau = E_2 F(\br), \eqno (19)
$$
where $\br = \bx - \by$,  and 
$\ds V(\br) = {\lambda \over {2m^2}} \delta^3(\br)$.

It is of interest to write out this equation in component form, with
$$
F(\br) = \left [ \matrix {s(\br) & t(\br)\cr u(\br) & v(\br)\cr} \right ], \eqno (20)
$$
namely
$$
-{1\over {2m}}\nabla^2(2s -u -t) + V (s-t-u+v) + (2m - E_2) s = 0, \eqno (21)
$$
$$
-{1\over {2m}}\nabla^2(s - v) + V (s-t-u+v) - E_2 t = 0, \eqno (22)
$$
$$
-{1\over {2m}}\nabla^2(s - v) + V (s-t-u+v) - E_2 u = 0, \eqno (23)
$$
and
$$
-{1\over {2m}}\nabla^2(t+u-2v) + V (s-t-u+v) - (2m + E_2) v = 0. \eqno (24)
$$
Equations (22) and (23) imply that $t(\br) = u(\br)$, so that only three equations survive:
$$
(2m - E_2 - \kem + V) s + (\kem  - 2 V)t + V v = 0, \eqno (25)
$$
$$
(-\kemm + V) s - (E_2 + 2 V) t + (\kemm + V) v = 0, \eqno (26) 
$$
and
$$
V s - (\kem + 2 V)t - (2m + E_2 - \kem - V) v = 0. \eqno (27)
$$
These equations have positive-energy solutions of the type $E_2 = m+m+\cdots$, negative-energy solutions of the type $E_2=-m-m+\cdots$, and ``mixed'' type solutions with $E_2=m-m+\cdots$ (this is clear, for example, if $V=0$ and the particles are at rest). 

For the positive-energy solutions, if we write $E_2=2m+\epsilon$, then in the non-relativistic limit $|(\epsilon - V + {\ds {p^2\over m}})v| \ll |m v|$ (and similarly for $s$ and $t$), and so equations (26) and (27) show that $t$ and $v$ are small components, by factors $O({\ds{\epsilon\over m}, {p^2\over {m^2}}, {V\over m}})$. Thereupon, equation (25) reduces to
$$
-\kem s(\br) + V(\br) s(\br) = \epsilon s(\br), \eqno (28)
$$
which is the usual time-independent Schr\"odinger equation for the relative motion of two particles, each of mass $m$, interacting through the potential $\ds V(\br) = {\lambda \over {2m^2}} \delta^3(\br)$. Similarly, in the non-relativistic limit, $v$ is the large component for the negative-energy solutions (with $E_2 = -(2m+\epsilon)$, and $s \to v$, $V \to -V$ in (28)), while $t$ is the large component for the mixed  energy solutions. This is obvious from the form of the free-particle solutions ($V=0$), which are
$$
F(\br) = s_0 \left [\matrix{  1& ({p\over{\omega+m}})^2\cr ({p\over{\omega+m}})^2& ({{\omega-m}\over{\omega+m}})^2\cr } \right ] e^{i\bp\cdot\bx} \;\;\;\;  
{\longrightarrow \atop {{p\over m} \ll 1}}
\;\;\;\; s_0 \left [ \matrix{  1 & ({p\over {2m}})^2   \cr ({p\over {2m}})^2 & ({p\over {2m}})^4  \cr  } \right ] e^{i\bp\cdot\bx}, \eqno (29)
$$
for $E_2 = 2 \omega = 2 \sqrt {p^2+m^2}$,
$$
F(\br) = t_0 \left [\matrix{{p^2\over{2m^2+p^2}}& 1 \cr 1 & {p^2\over{2m^2+p^2}}\cr}\right ]e^{i\bp\cdot\bx}, \eqno (30)
$$
for $E_2 = 0$, and
$$
F(\br) = v_0 \left [\matrix{ ({{\omega-m}\over{\omega+m}})^2& ({p\over{\omega+m}})^2\cr ({p\over{\omega+m}})^2& 1\cr}\right ]e^{i\bp\cdot\bx}\;\;\;\;
{\longrightarrow \atop {{p\over m} \ll 1}}
 \;\;\;\; v_0 \left [\matrix{({p\over {2m}})^4& ({p\over {2m}})^2\cr ({p\over {2m}})^2& 1\cr}\right ]e^{i\bp\cdot\bx}, \eqno (31)
$$
for $E_2 = - 2 \omega = - 2 \sqrt {p^2+m^2}$, and where $s_0, t_0$ and $v_0$ are constants.

For the repulsive delta-function potential there are, of course, no bound state solutions.  For the continuum case, the scattering is trivial, in that the phase-shifts are zero (the $S$-matrix is unity), as 
we point out in detail below.
 Thus the interaction is ``trivial'' in this sense. 
B\'eg and Furlong [4] have shown previously that the non-relativistic limit of $\lambda (\varphi^* \varphi)$ theory corresponds to the same 'trivial' repulsive delta-function interactions. The present results extend this to the relativistic case.
 These results are also consistent with the generally accepted ``triviality'' of $\lambda\,\varphi^4$ theory, in the sense that the particle excitations above the vacuum are non-interacting (e.g. references [5] and [6], and citations therein).

We now proceed to demonstrate the triviality of the scattering by explicit solution.  Equations (25)-(27) can be reduced by taking suitable linear combinations, whereupon it follows that ($E = E_2$)
$$
(2m +E)v = (2m-E)s + 2 E t \eqno (32)
$$
and
$$
[E(4m^2-E^2) - 8 m^2 V] s = - [E(2m+E)^2 + 8 m^2 V]t . \eqno (33)$$
It is easily verified that the free particle solutions (29) - (31), in particular, satisfy these relations.  One can therefore write
$$
s = [8m^2V+E(2m+E)^2]\chi \eqno (34)
$$ 
and
$$
t = [8m^2V+E(E^2-4m^2)]\chi, \eqno (35)
$$
where $\chi$ is a solution of
$$
-4E \nabla^2 \chi + [E(4m^2-E^2) + 8m^2V]\chi = 0. \eqno (36)
$$
Equation (36) is form-identical to the Schr\"odinger equation. Since the delta-function interaction becomes simply  a constant in the momentum representation, it is convenient to write (36) in momentum space:
$$
(p^2-\kappa^2) \chi(\bp) + {\lambda \over {(2\pi)^3E}}\int d^3q \,\chi(\bq) = 0, \eqno (37)
$$
where $\kappa^2 = (E/2)^2-m^2$, and from which it is obvious that only s-waves are affected by the delta-function potential. Thus, we can write the solution of (37) as
$$
\chi(p) = {1\over{4\pi p^2}} \left [ \delta(p-\kappa) + {{\lambda\, c_{\chi}\, p^2} \over {2\pi^2 E (p^2-\kappa^2)}}\right ]\;, \eqno (38)
$$
where
$$
c_{\chi} = \int d^3q\, \chi (\bq). \eqno (39)
$$
The $s$-wave phase shift, extracted from eq. (38), is then given by
$$
tan \,\eta = {{\lambda\, c_{\chi}\, \kappa} \over {4\pi E}}. \eqno (40)
$$
Substituting eq. (38) into eq. (39) and solving for $c_{\chi}$ yields the result
$$
c_{\chi} = {1 \over {\ds{{1 - {{\lambda I_1}\over {2\pi^2 E}}}}}}, \eqno (41)
$$
where
$$
I_1 = {\cal P} \int_0^{\Lambda} dp\, p^2\, {1\over {p^2 -\kappa^2}}. \eqno (42)
$$
In actuality, the upper limit $\Lambda$ on the integral (42) is infinite, however, the integral then diverges linearly with $\Lambda$, hence we regulate it with this cut-off. Nevertheless, when finally we take $\Lambda \to \infty$, equation (41) shows that $c_{\chi} \to 0$, hence the phase-shift $\eta$ of eq. (40) vanishes for any finite value of $\lambda$. This confirms that the $S$-matrix is indeed unity.

One could, of course, solve equations (25)-(27) directly (and so eq. (19)) since, for a delta-function potential, they are, in essence, algebraic equations in momentum space. This leads to the same results, albeit at the cost of a little more algebraic effort.

It is straightforward to write down three-body eigenstates analogous to (16), namely
$$
|\psi_3\rangle = \int d^3x_1\, d^3x_2\,d^3x_3\;F_{abc}(\bx_1,\bx_2,\bx_3)\,\phi_a^{\dag}(\bx_1) \phi_b^{\dag}(\bx_2) \phi_c^{\dag}(\bx_3)|\tilde 0\rangle, \eqno (43)
$$
provided that the $2^3=8$ coefficient functions $F_{abc}(\bx_1,\bx_2,\bx_3)$ are solutions of the relativistic three-body KG-FV-like equation 
$$
\eqalignno{
{\tilde h}&_{ak}(\bx_1) F_{kbc}(\bx_1,\bx_2,\bx_3) +
{\tilde h}_{bk}(\bx_2) F_{akc}(\bx_1,\bx_2,\bx_3) +
{\tilde h}_{ck}(\bx_3) F_{abk}(\bx_1,\bx_2,\bx_3)\cr
&+ V(\bx_1-\bx_2) {\tilde \tau}_{ak_1} {\tilde \tau}_{bk_2} F_{k_1k_2c}(\bx_1,\bx_2,\bx_3)
 + V(\bx_2-\bx_3) {\tilde \tau}_{bk_1} {\tilde \tau}_{ck_2} F_{ak_1k_2}(\bx_1,\bx_2,\bx_3) & (44)\cr
&+ V(\bx_3-\bx_1) {\tilde \tau}_{ck_1} {\tilde \tau}_{ak_2} F_{k_1bk_2}(\bx_1,\bx_2,\bx_3)
= E_3 F_{abc}(\bx_1,\bx_2,\bx_3),\cr}
$$  
where summation on repeated indices is implied and where $\ds V(\bx_i-\bx_j) = {\lambda \over {2m^2}} \delta^3(\bx_i-\bx_j)$. Once again we have ``trivial'' delta-function (contact) interactions among the particles, exactly as in the two-body case.  Generalizations for $N$-body eigenstates can be written down in an analogous fashion.

I would like to thank P. M. Stevenson and M. Horbatsch for interesting conversation and helpful suggestions.
The financial support of the Natural Sciences and Engineering Research Council of Canada for this work is gratefully acknowledged.

\vfill \eject
{\ni \bf References}
\medskip
\par
\item{1.} F. Gross, {\sl Relativistic Quantum Mechanics and Field Theory} (Wiley-Interscience, New York, 1993).
\par\noindent
\item{2.} H. Feshbach and F. Villars, Rev. Mod. Phys. {\bf 30}, 24 (1958).
\par\noindent
\item{3.} J. D. Bjorken and S. D. Drell, {\sl Relativistic Quantum Fields} (McGraw Hill, New York, 1965).
\par\noindent

\item{4.} M. A. B. B\'eg and R. C. Furlong, Phys. Rev. D {\bf 31}, 1370 (1985). 

\item{5.} M. Consoli and P. M. Stevenson, Z. Phys. C {\bf 63}, 427 (1994).
\par\noindent

\item{6.} M. Consoli and P. M. Stevenson, {\sl Resolution of the $\lambda \phi^4$ puzzle}, Rice University preprint, July 1993, DE-FG05-92ER40717-5 (hep - ph 930 3256).
\par\noindent  

\bye